\documentclass[twocolumn,showpacs,aps,prd,nobibnotes,floatfix]{revtex4-1}
\newcommand{\beq}{\begin{equation}}
\newcommand{\eeq}{\end{equation}}
\newcommand{\beqn}{\begin{eqnarray}}
\newcommand{\eeqn}{\end{eqnarray}}

\newcommand{\apjl}{Astrophys.\ J.\ Lett.}

\newcommand{\mnras}{Mon. Not. Roy. Astron. Soc.}

\newcommand{\sunmass}{M_{\odot}}

\usepackage{graphicx}
\usepackage{epsf}
\usepackage{graphics,epsfig,placeins,subfigure,wrapfig}
\begin{document}
\title{General relativistic simulations of black hole-neutron
  star mergers: \\ Effects of tilted magnetic fields}

\author{Zachariah B. Etienne${}^1$, Vasileios Paschalidis${}^1$, and Stuart L. Shapiro${}^{1,2}$
}                           
%
\affiliation{
${}^1$ Department of Physics, University of Illinois at Urbana-Champaign,
Urbana, IL 61801 \\
${}^2$ Department of Astronomy and NCSA, University of Illinois at Urbana-Champaign,
Urbana, IL 61801 }


\begin{abstract}
Black hole--neutron star (BHNS) binary mergers can form disks in
which magnetorotational instability (MRI)-induced turbulence may drive
accretion onto the remnant BH, supporting relativistic jets and providing the
engine for a short-hard gamma-ray burst (SGRB). Our earlier study of
magnetized BHNSs showed that NS tidal disruption winds the magnetic
field into a toroidal configuration, with poloidal fields so weak
that capturing MRI with full-disk simulations would require $\sim 10^8$
CPU-hours. In that study we imposed equatorial symmetry, suppressing
poloidal magnetic fields that might be generated from plasma crossing
the orbital plane. Here we show that initial conditions
that break this symmetry (i.e., {\it tilted} poloidal magnetic fields
in the NS) generate much stronger poloidal fields in
the disk, indicating that asymmetric initial conditions may be
necessary for establishing BHNS mergers as SGRB progenitors via
fully general relativistic MHD simulations. 
We demonstrate that BHNS mergers may form an SGRB
engine under the right conditions by seeding the remnant disk from
an unmagnetized BHNS simulation with purely poloidal fields
dynamically unimportant initially, but strong
enough to resolve MRI. Magnetic turbulence occurs in the
disk, driving accretion and supporting Poynting-dominated jet outflows
sufficient to power an SGRB. 
\end{abstract}

\pacs{04.25.D-,04.25.dk,04.40.Nr}

\maketitle

\section{Introduction}
At the end of a black hole--neutron star (BHNS)
binary merger, the NS may tidally disrupt, forming a hot, massive
disk around the BH, with the temperature, density, and collimated
magnetic field profiles required to launch a jet and trigger a short-hard
gamma-ray burst (SGRB). 

Recently a suite of stationary BH+disk general relativistic MHD (GRMHD) simulations was performed,
in which the disks were
seeded with different magnetic field topologies and
strengths~\cite{GRMHD_Jets_Req_Strong_Pol_fields}.
It was found that to support a long-term, Poynting-dominated jet,
sufficiently strong dipole poloidal fields are required in the
disk initially. 

This result may be problematic for BHNS mergers,
because the NS tidal disruption event invariably winds the magnetic fields
into a toroidal configuration in all magnetized BHNS simulations to
date. For example, in~\cite{UIUC_MAGNETIZED_BHNS_PAPER1}, we performed
the first parameter survey of magnetized BHNS mergers using fully
general relativistic MHD
simulations, varying the initial magnetic field strength and geometry,
and the aligned BH spin. We seeded the NS with purely poloidal
magnetic fields and found that after tidal disruption, the fluid
motion becomes strongly toroidal during disk formation, dragging the
magnetic field lines into a {\it predominantly toroidal}
configuration, leaving very weak poloidal fields.

However, even these very weak poloidal fields can be exponentially
amplified via the magnetorotational instability (MRI). But the weaker
the poloidal fields, the shorter the wavelength of the fastest-growing
MRI mode. Numerically, if this wavelength is not resolved by
at least 10 gridpoints, MRI cannot be
captured numerically~\cite{MRI_RESOLVED_TO_10_GRIDPOINTS_PAPER}. We performed a
local analysis of our magnetized BHNS disks formed
in~\cite{UIUC_MAGNETIZED_BHNS_PAPER1} and found that the
fastest-growing MRI wavelength was under-resolved by about a factor
of 10 on average. These simulations required about 30,000 CPU-hours and
two weeks of wall-clock time, so resolving MRI would require about
$3 \times 10^8$ CPU-hours and 20 weeks of wall-clock time, which is
computationally unfeasible.

Resolving MRI is considerably more challenging in BHNS simulations
than in typical magnetized NSNS simulations
\cite{PR2006,AHLLMNPT2008,LSET2008,BrunoMagNSNS,ML2011}.
First, in NSNS mergers, the Kelvin-Helmholtz instability amplifies the
poloidal magnetic field by an order of magnitude when the stars come
in contact~\cite{PR2006,AHLLMNPT2008}.  Second, in
BHNS mergers, the NS core containing most of the magnetic energy is
quickly swallowed by the BH following tidal disruption, leaving only
weakly magnetized matter in the outer layers to form a disk which is
more extended than disks formed in NSNS mergers.  Third,
frame-dragging in BHNS mergers enhances the winding of the magnetic
field, reducing the poloidal component even further. The net effect is
that the wavelength of the fastest-growing MRI mode is much smaller in
merging BHNSs than NSNSs with the same initial magnetized NS.

In previous magnetized BHNS
simulations~\cite{UIUC_MAGNETIZED_BHNS_PAPER1,Chawla:2010sw}, weak
poloidal fields in the remnant disk made it impossible to resolve
MRI-induced turbulence, which would drive accretion, and possibly
launch and sustain jets. Additionally, in our earlier study the system
obeyed equatorial symmetry, which saved computational cost. However,
these symmetric configurations prevent plasma motion across the
orbital plane and thus poloidal magnetic fields that might be
generated from such motion. Therefore we hypothesize that {\it
introducing asymmetries in this system will very likely enhance the
poloidal fields in the remnant disk, possibly enabling us to capture
MRI}.  Such asymmetries may be common in nature, given the magnetic
field misalignment of the double pulsar system PSR
J0737-3039~\cite{DoublePulsar}.

Here we seed the pre-disrupted NS with the same purely
poloidal fields as our previous work, but also {\it break the
  symmetry} by tilting them relative to the orbital angular momentum.
We find that as the tilt angle increases to 90$^{\circ}$, poloidal
fields in our remnant disk become stronger, confirming our
hypothesis. But the resulting field strengths are not sufficient to
resolve MRI, establish turbulence, or launch a jet.

The question then arises: could the presence of a dominant poloidal
dipole field in the remnant BHNS disk produce an SGRB engine? To
answer this, we seed the remnant disk from an unmagnetized BHNS
simulation with a purely dipole poloidal field, dynamically
unimportant initially, but strong enough to resolve MRI. Within a few
orbital periods, turbulence sets in and above the BH poles the
magnetic fields collimate and relativistic outflows turn on,
sufficient to power an SGRB. This is the first GRMHD simulation to
demonstrate that jets can be produced from a disk formed at the end of
a BHNS merger for a suitable remnant field.

We conclude that strong dipole poloidal fields in the remnant disks
from BHNS mergers can give rise to MRI and launch
Poynting-dominated outflows. Forming such poloidal fields strong
enough to resolve MRI self-consistently through GRMHD simulations will
likely require asymmetric initial conditions.

The paper is structured as follows. In Sec.~\ref{sec:basic_eqns}
we describe the initial data, basic evolution equations, and 
numerical methods. The basic results are presented in Sec.~\ref{sec:results} 
and summarized in Sec. \ref{sec:summaryandfuturework}.
Throughout this work, geometrized units are
adopted, where $G = c = 1$, unless otherwise specified.

\section{Basic Technique}
\label{sec:basic_eqns}

Our simulation software and methods are the same as used for case B4
in~\cite{UIUC_MAGNETIZED_BHNS_PAPER1}, except for a few key
improvements. First, in our vector potential formulation for the
induction equation, we adopt the ``Generalized Lorenz gauge'' (GL) we
developed in~\cite{BriansLatest2012}, choosing $1/\xi=0.45M$. This
modification results in {\it damped}, traveling EM gauge modes,
preventing spurious magnetic fields from appearing on refinement boundaries
more effectively than our original, undamped Lorenz gauge
condition~\cite{UIUC_MAGNETIZED_BHNS_PAPER1}.  Second, we fixed a
coding error introduced in our attempt to make the primitives solver
more robust in~\cite{UIUC_MAGNETIZED_BHNS_PAPER1}; this error affected only
that study.  The net effect of this error was a doubling of
truncation errors, but ultimately it had no effect on the bulk dynamics of
our simulations. After fixing the error, for example, the small disk mass
amplification and gravitational wave mismatch found in our previous
work when comparing unmagnetized cases (A0 and B0) to the strongest
magnetic field cases (A4 and B4)~\cite{UIUC_MAGNETIZED_BHNS_PAPER1}
are significantly reduced.

Finally, we no longer impose symmetry across the orbital plane, except
in one of our simulations.  Not imposing symmetry across the orbital
plane enables us to study asymmetric magnetic field configurations.
In this work, we apply a rotation matrix to the vector potential
[Eqs. (11) \& (12) of~\cite{UIUC_MAGNETIZED_BHNS_PAPER1}] to tilt the
purely poloidal magnetic fields 45$^{\circ}$ and $90^{\circ}$ relative
to the orbital angular momentum vector.

We have also added two new diagnostics to better monitor magnetic
effects. The first diagnostic monitors the fastest-growing MRI
wavelength~\cite{FASTEST_GROWING_MRI_WAVELENGTH},

\beq
\label{lambdaMRI}
\lambda_{\rm MRI} \approx 2 \pi \frac{|v_{\theta,A}|}{|\Omega(r,\theta)|}
\approx 2 \pi \frac{\sqrt{|b_{P}b^{P}|/(b^2+\rho_0 h)}}{|\Omega(r,\theta)|},
\eeq
where $|b^{P}| \equiv \sqrt{b_\mu b^\mu - |b_\mu(e_{\hat{\phi}})^\mu|^2}$, and $(e_{\hat{\phi}})^{\mu}$ is the toroidal orthonormal vector
carried by an observer comoving with the fluid.
The second new diagnostic measures Poynting flux across a surface
$\mathcal{S}$, $L_{\rm EM} \equiv -\int T^{r (\rm EM)}_{t} \sqrt{-g}
dS$, where $T^{\mu (\rm EM)}_{\nu}$ is the electromagnetic stress-energy tensor.

\section{Results}
\label{sec:results}

\begin{figure*}
\epsfxsize=3.2in
\leavevmode
\epsffile{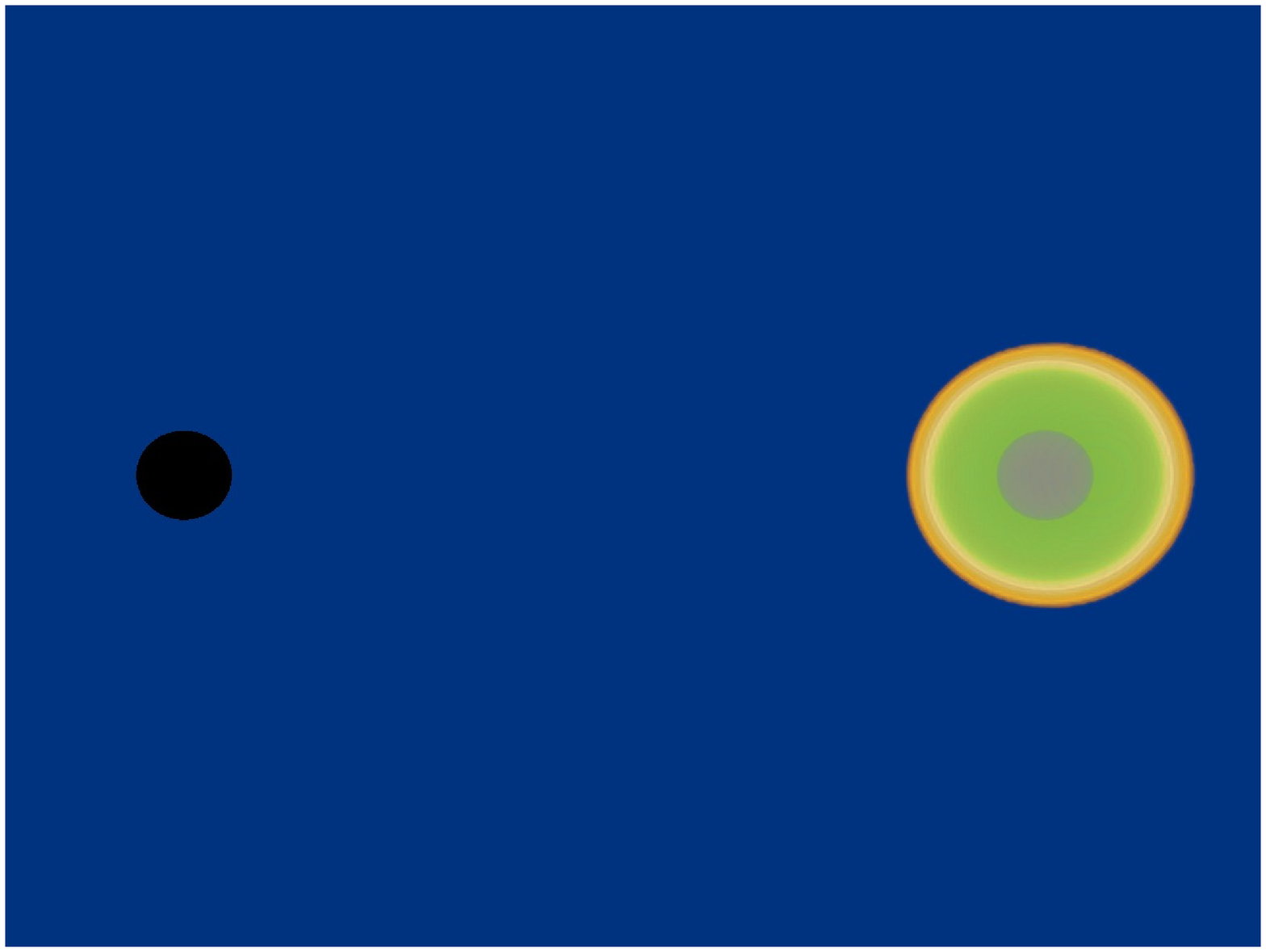}
\epsfxsize=3.2in
\leavevmode
\epsffile{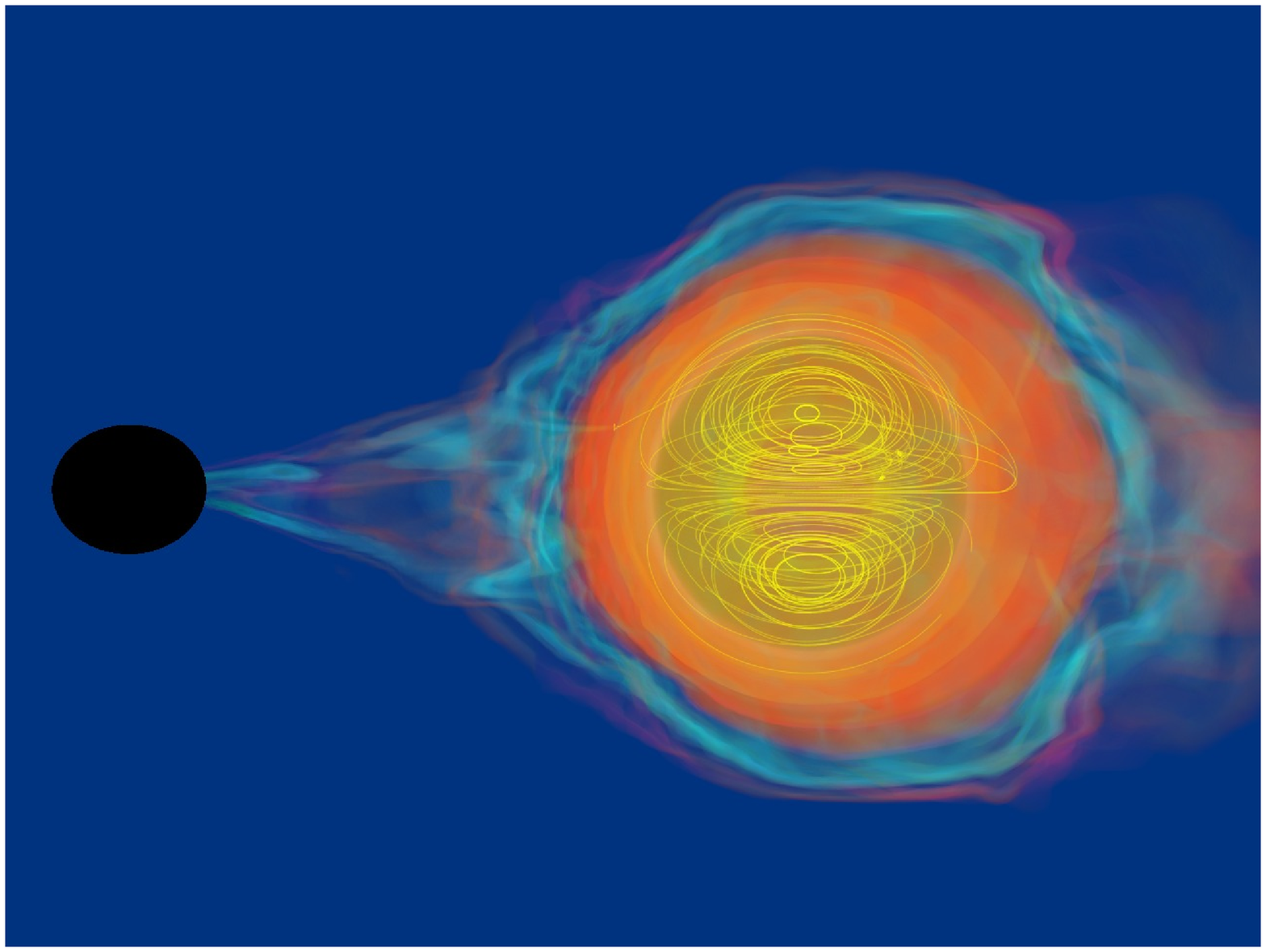} \\
\epsfxsize=3.2in
\leavevmode
\epsffile{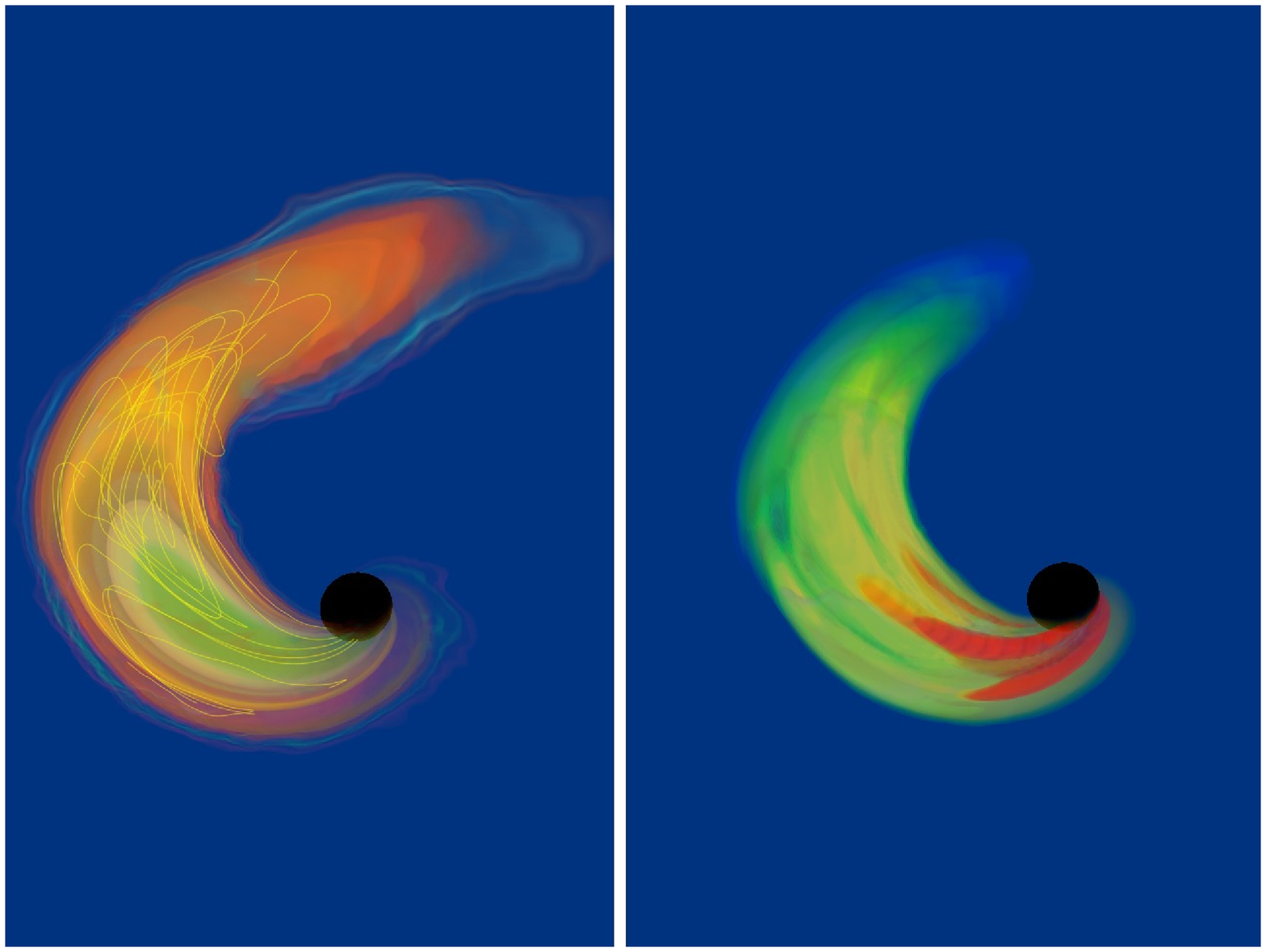}
\epsfxsize=3.2in
\leavevmode
\epsffile{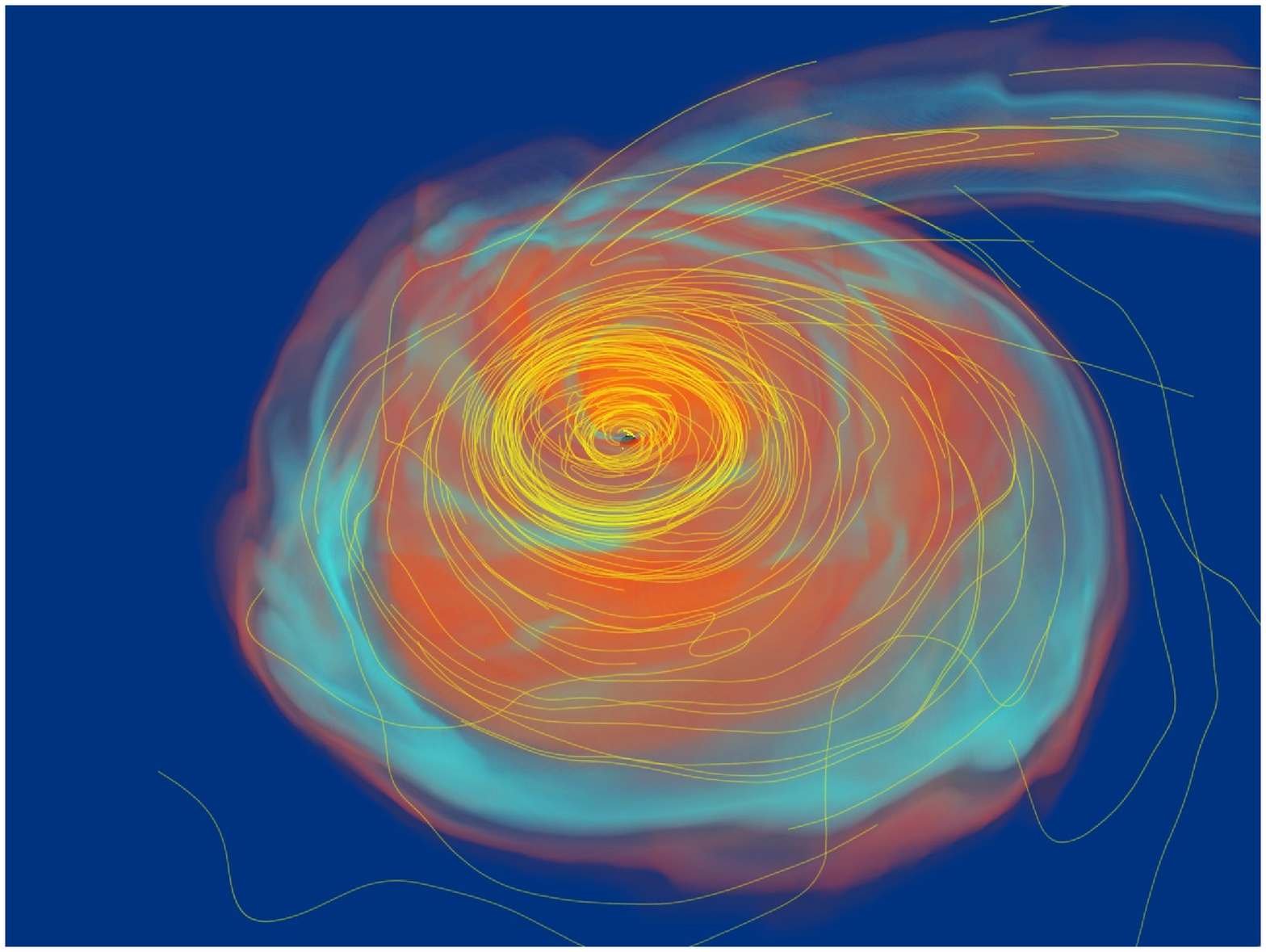}
\caption{Case B4-90 3D snapshots. Top-left: initial data, NS on
  the right (from highest to lowest rest-mass density, the colors are:
  grey, green, yellow, orange, red, and blue), BH apparent horizon (AH) on the
  left. Top-right: magnetic fields (yellow) are seeded into NS after
  $\approx 4$ orbits (simulation time $t=633M$). Bottom-left:
  tidal disruption after $\approx 5$ orbits ($t=882M$); density and magnetic
  field lines on the left, and magnetic energy density $b^2/2$ on the
  right (from highest to lowest $b^2/2$, the colors are: red,
  yellow, and green). Bottom-right: final disk density profile with magnetic field
  lines, about $33$ms $(1.4 \sunmass/M_0)$ after disk
  formation ($t=2072M$), where $M_0$ is the initial rest mass of the NS. 
}
\label{figbasicstory}
\end{figure*}

Our initial data are generated using the conformal thin-sandwich (CTS)
formulation~\cite{eflstb08,UIUC_BHNS__BH_SPIN_PAPER,baumgartebook10},
with the NS modeled as an irrotational, unmagnetized, $n = 1$
polytrope. The initial binary is in a quasi-equilibrium circular orbit
just outside the tidal disruption radius, with the BH irreducible mass
three times that of the isolated NS ADM mass.  The BH spin parameter
is set to $a/M=0.75$, aligned with the orbital angular momentum
(top-left frame of Fig.~\ref{figbasicstory}). As
in~\cite{UIUC_MAGNETIZED_BHNS_PAPER1}, the BH and NS in our
simulations are covered by 70 and 80 gridpoints on average across
their shortest diameters, respectively. Our grid consists of 8 and 9
refinement boxes centered on the NS and BH, respectively, with outer
boundary at $210M$.

We first evolve a total of five cases: B0, B4-0, B4-45, B4-90, and
B4-0-Sym. B4-0-Sym, which imposes symmetry across the orbital plane,
is identical to case B4 in~\cite{UIUC_MAGNETIZED_BHNS_PAPER1}, but is
now evolved with our latest code. B4-0 is the same as B4-0-Sym but
does not impose equatorial symmetry. B4-45 and B4-90 are the same as
B4-0, but the initial poloidal NS magnetic fields are tilted by
$45^{\circ}$ and $90^{\circ}$, respectively. Finally, B0 is the same
as B4-0, except with zero magnetic fields.

We evolve the unmagnetized NS for nearly four orbits before seeding
its interior with a purely poloidal, $\sim 10^{16}$G magnetic field,
in the magnetized cases. With average magnetic-to-gas pressure of
0.5\%, these magnetic fields are dynamically unimportant to the
NS. The second panel of Fig.~\ref{figbasicstory} shows B4-90 when the
magnetic fields are first seeded in the NS. Though a NS cannot stably
maintain purely poloidal
fields~\cite{Stable_NS_cannot_have_poloidal_fields1,Stable_NS_cannot_have_poloidal_fields2},
we choose such an initial configuration, together with a very high
field amplitude, to maximize the prospect of obtaining large residual
poloidal fields in the remnant disk, making it more computationally
feasible to resolve the MRI (cf. Eq.~\ref{lambdaMRI}).

\begin{figure}
\vspace{0.75cm}
\epsfxsize=3.2in
\leavevmode
\epsffile{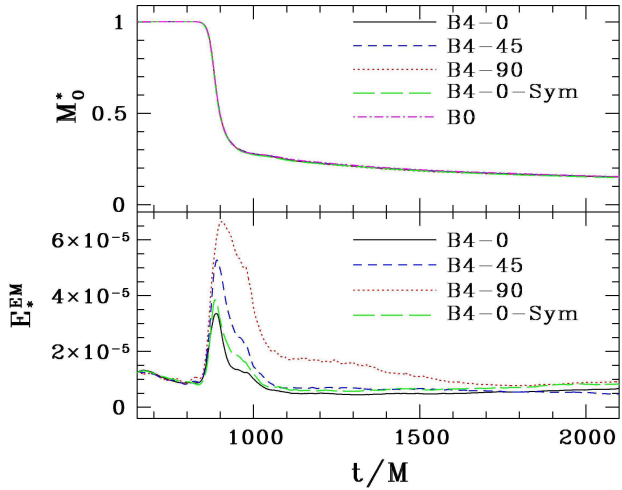}
\caption{Rest mass outside the AH $M^*_0$ (top frame) and
  magnetic energy outside the AH $E^{\rm EM}_*$ (bottom frame),
  versus time. Rest mass and magnetic energy are given by Eqs.~(23) and (13)
  in~\cite{UIUC_MAGNETIZED_BHNS_PAPER1}, respectively.}
\label{mag_energy}
\end{figure}

The amplitude of the magnetic field seeds in the NS scales like the
local gas pressure and is thus strongly peaked at the core of the
NS. During tidal disruption, the core of the NS is accreted
(bottom-left frame in Fig.~\ref{figbasicstory}), which would seem to
leave very little magnetic energy available for the remnant
disk. However, the tidal disruption of the orbiting NS stretches its
frozen-in magnetic fields, leading to a net {\it amplification} of
magnetic energy. The bottom frame of Fig.~\ref{mag_energy}
demonstrates that midway through tidal disruption (accretion history
is plotted in the top frame) the magnetic energy outside the AH peaks
at $\sim 2$ times the initial value, then rapidly drops, so that the
final disk has roughly the same magnetic energy as the initial seed
magnetic fields.

After evolving the disks in cases B4-0, B4-45, B4-90, and B4-0-Sym for
about $1200M$, $14$\% of the NS rest mass remains outside the AH,
regardless of the presence or tilt of the magnetic field (see top
panel of Fig.~\ref{mag_energy}).  If this mass were converted into
gamma-rays at an efficiency of 1\%, the energy output would be of
order $10^{51}$ergs, sufficient to launch an SGRB. However, for an
SGRB to take place, models indicate that jets must be launched from
the final disk, and jet-launching may fail to occur if the disk
magnetic fields are not sufficiently dipolar and poloidal near the
BH~\cite{GRMHD_Jets_Req_Strong_Pol_fields}.

When we terminate B4-90, the disk magnetic fields are almost purely
{\it toroidal} (lower-right frame of Fig.~\ref{figbasicstory}), and
fluid {\it inflow} continues in the BH polar regions. MRI could
amplify the remnant poloidal fields, but only if $\lambda_{\rm
  MRI}/\Delta x\gtrsim 10$, where $\Delta x$ is the local
gridspacing~\cite{MRI_RESOLVED_TO_10_GRIDPOINTS_PAPER}.  On average,
this ratio is only $\approx 1$ in the B4-0 and B4-0-Sym disks, as
these cases are by construction equatorially symmetric. The ratio
increases to $\approx 3$ in B4-45 and to $\approx 8$ in
B4-90. Computational cost goes like $(\Delta x)^{-4}$ in these
simulations, so MRI can be resolved in case B4-90 at about
$(\frac{1}{10})^{-4}/(\frac{8}{10})^{-4}=1/4096$ the cost of B4-0.
However, even though we are using high-resolution grids and very
strong initial magnetic fields, we are still not able to capture MRI
and its effects.

\begin{figure*}
\vspace{-4mm}
\epsfxsize=3.2in
\leavevmode
\epsffile{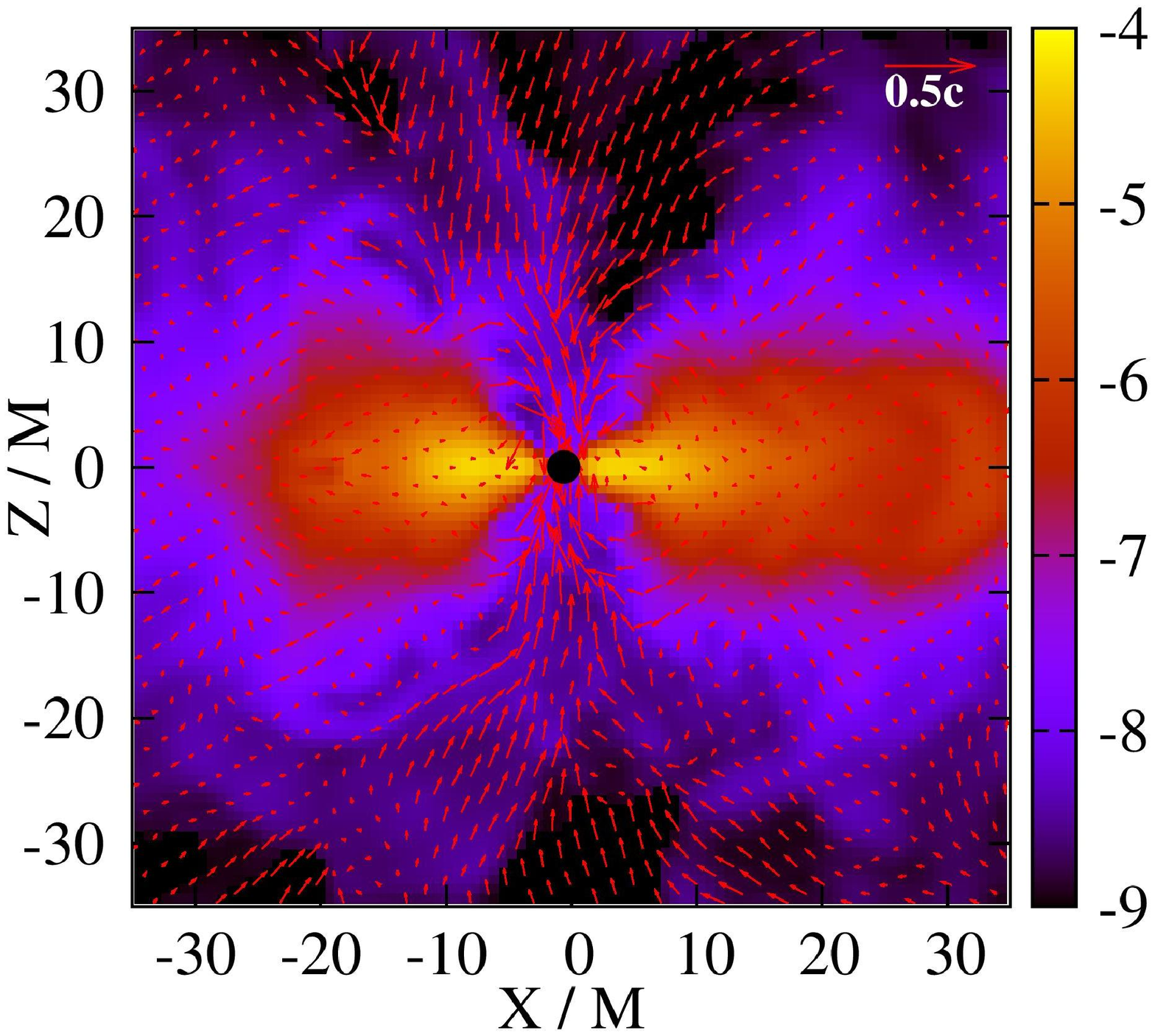}
\epsfxsize=3.2in
\leavevmode
\epsffile{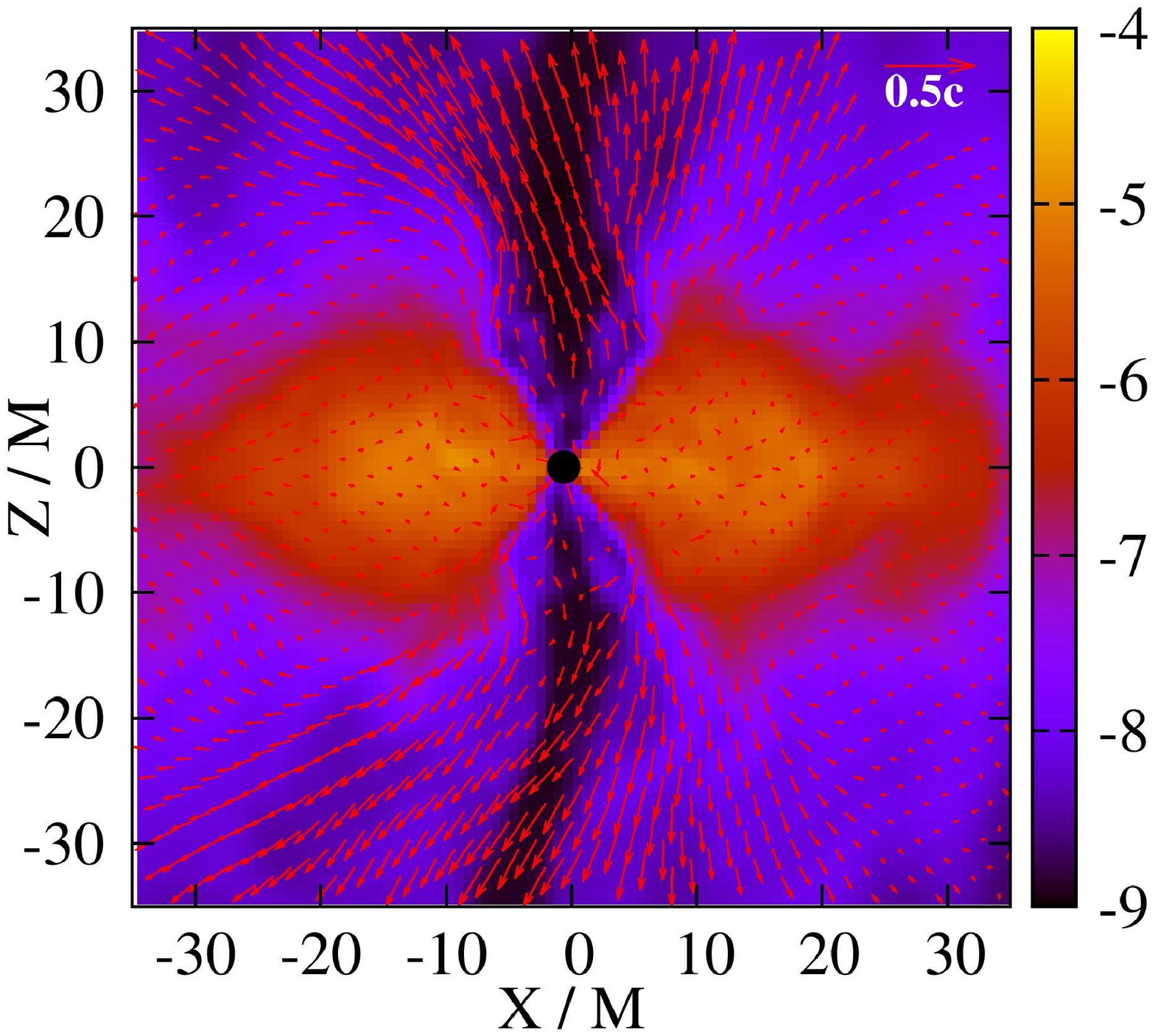} \\
\epsfxsize=3.2in
\leavevmode
\epsffile{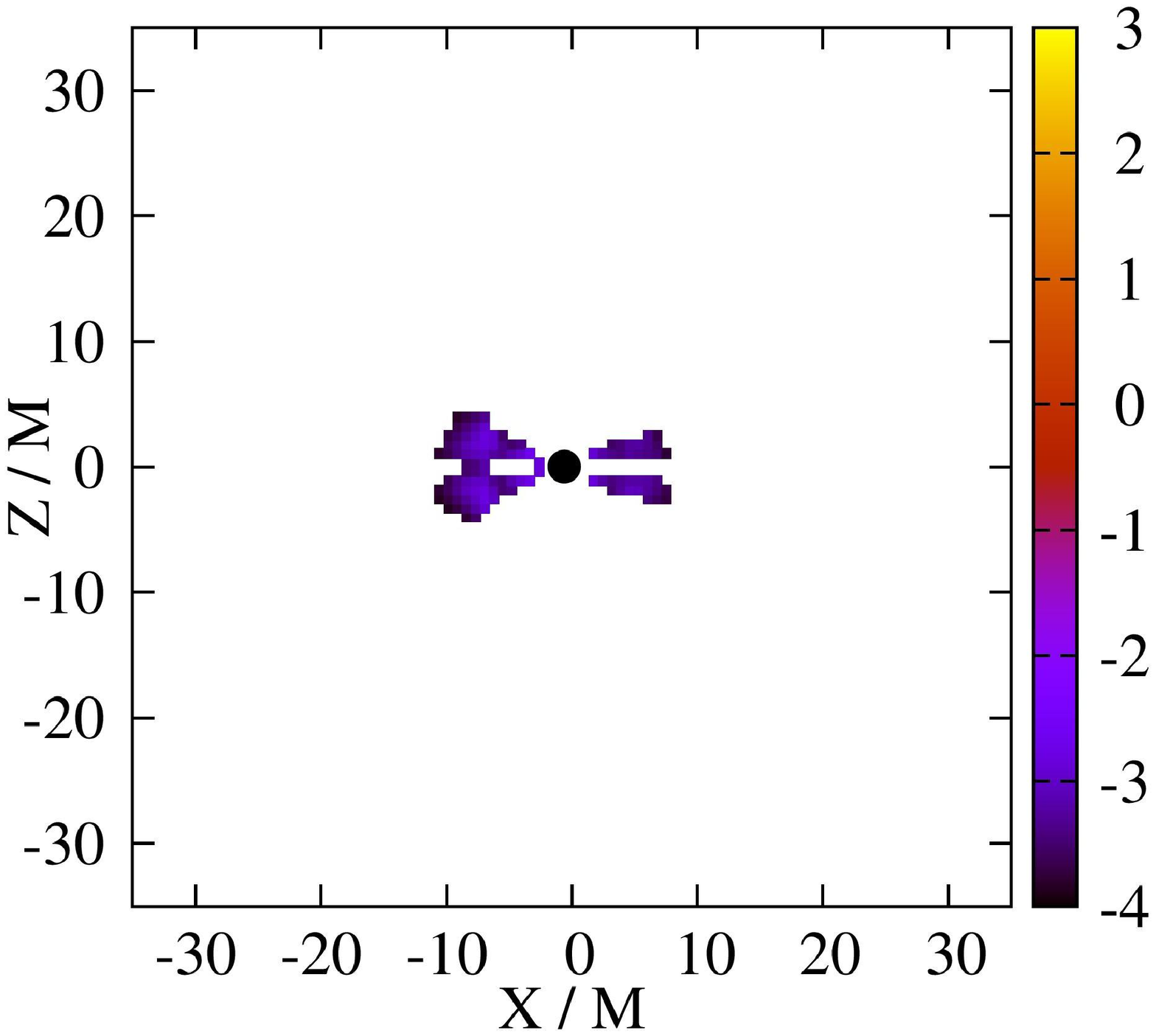}
\epsfxsize=3.2in
\leavevmode
\epsffile{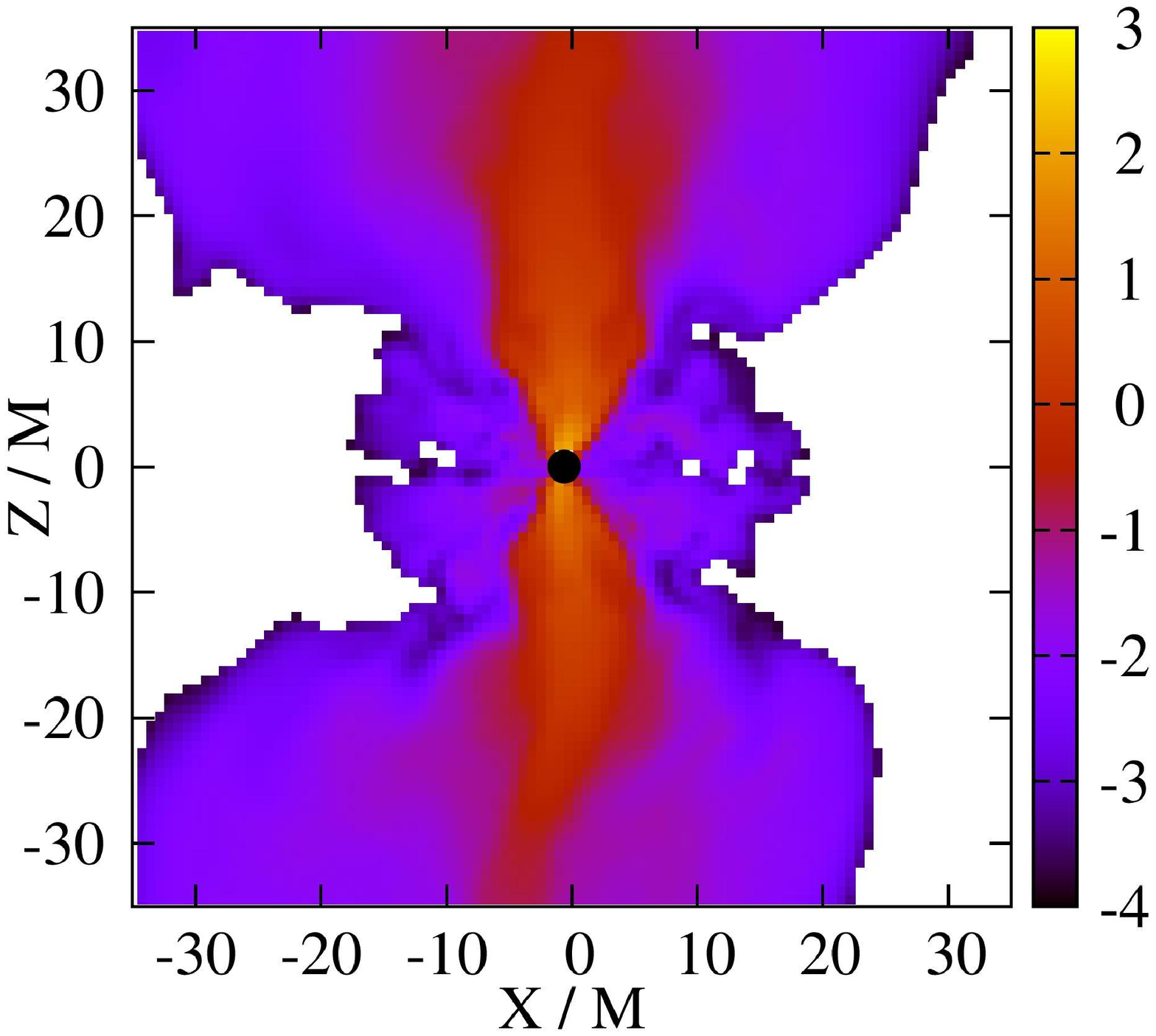} \\
\epsfxsize=3.2in
\leavevmode
\epsffile{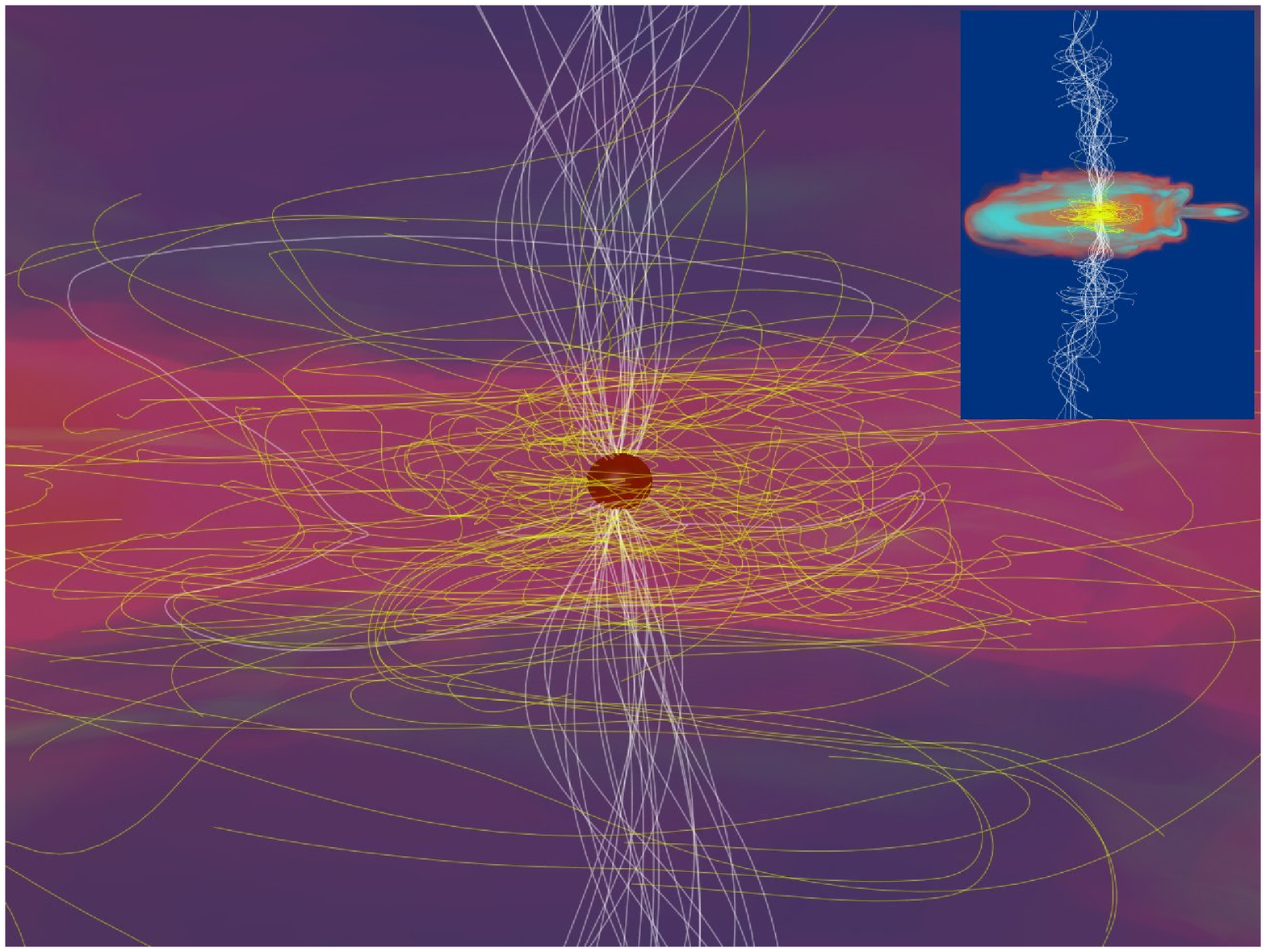}
\epsfxsize=3.2in
\leavevmode
\epsffile{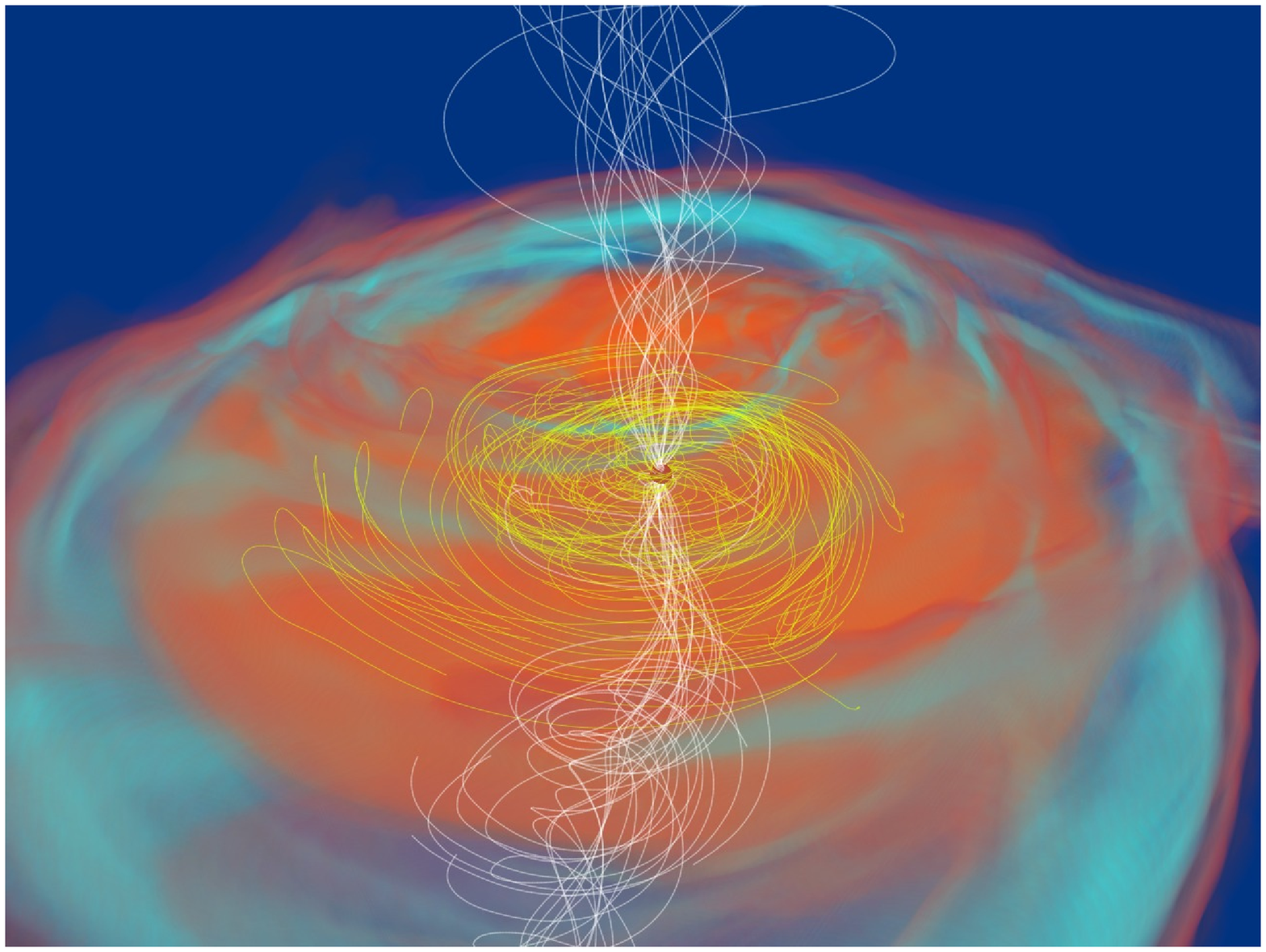}
\caption{Case B5 snapshots. Top row: map of $\log_{10} \rho_0$ on the
  meridional plane that passes through the BH centroid, at the time
  when the disk is seeded ($t=2152M$) with magnetic fields (left frame) and
  when we terminated the simulation at $t=2598M$ (right frame). Velocity
  arrows projected onto this plane are also displayed (red arrows). The
  initial central density of the NS is $9\times
  10^{14} \mbox{g cm}^{-3} (1.4M_\odot/M_0)^2$ (0.13 on the color bar), where $M_0$ is the
  initial NS rest mass. Middle row: map of $\log_{10} b^2/(2\rho_0)$ on the
  same meridional slices and times as the frames directly above
  them. Bottom row: 3D snapshots when we terminate the
  simulation, with viewing angles in the disk plane, both zoomed-in
  (left frame) and zoomed-out (left frame inset), and above the disk
  plane (right frame). Magnetic field streamlines emerging just above
  and below the BH poles are shown in white, and those in the disk are shown
  in yellow.}
\label{meridionaldiskB5}
\end{figure*}

To demonstrate that BHNS mergers may form an SGRB engine under the
{\it right} conditions, we evolve case B5, where we artificially seed
case B0's final disk with a purely poloidal magnetic field having an
average magnetic-to-gas pressure ratio of $\sim 0.1\%$, corresponding
to average magnetic field strength of $\approx 10^{14}(1.4M_\odot/M_0)
G$ (cf. B4-90's final disk, with $\approx
4\times10^{14}(1.4M_\odot/M_0)G$ average magnetic field
strength). Though these magnetic fields are dynamically weak
initially, they are strong enough to resolve MRI throughout most of
the disk because they are purely poloidal.  There are clear {\it
  inflows} above the BH poles when the disk is first seeded with
magnetic fields (upper-left panel of
Fig.~\ref{meridionaldiskB5}). Within $\sim 0.5$ orbital periods
magnetic turbulence begins, accretion of the dipole poloidal fields
occurs, and the inflows diminish.

Though $b^2/(2\rho_0)$ is at most only $\sim 10^{-4}$ initially
(Fig.~\ref{meridionaldiskB5}, middle-left panel), after $\sim 3$
orbital periods, it amplifies to a maximum value of $\gtrsim 100$ in
the Poynting-dominated funnels above and below the BH (middle-right
panel), and mildly-relativistic {\it outflows} appear above and below
the BH. For steady-state, Poynting-dominated jets, the energy-to-mass
flux ratio [which can be shown to be $\approx b^2/(2\rho_0)$] is equal
to the maximum possible Lorentz factor in the (asymptotic)
jet~\cite{B2_over_2RHO_yields_target_Lorentz_factor}. However, in
general, the actual terminal Lorentz factor of the jet is reduced by
the (neutrino-induced) baryon-loading into the jet funnel
\cite{Levinson2006,2007ApJ...659..561M,2011arXiv1112.2622L}.
But, the Blandford-Znajek process \cite{BZeffect}, which is captured
in our simulations, alone can accelerate a Poynting-dominated jet to
the necessary Lorentz factors even in the presence of baryon loading
\cite{2005astro.ph..6368M}. In fact, magnetic launching of
ultrarelativistic polar outflows from a BH-disk system is possible
even for nonspinning BHs provided the neutrino luminosity $L_\nu
\lesssim 10^{52}$ergs/sec \cite{Levinson2006}. Newtonian simulations of (10:1 mass-ratio)
BHNS mergers find characteristic neutrino
luminosities $L_\nu \sim 10^{51}$ergs/sec \cite{2005ApJ...634.1202R}. 
Moreover, neutron-rich outflows may result naturally in a high proton Lorentz factor
\cite{2006ApJ...650..998R}.  Thus, we expect that jets formed
following BHNS mergers can attain high Lorentz factors even when
accounting for baryon-loading in the jet.

When we terminate the B5 simulation, the Poynting luminosity is
$3.5\times10^{-3}\dot{M_0} c^2 =1.28\times10^{52}$ergs/sec, which may
be sufficient to power an SGRB. We terminated the simulation when
densities in the funnel region began to drop below our atmospheric
density. At the time of termination, $b^2/(2\rho_0)$ and the Poynting
luminosity were still increasing. The final magnetic fields are
turbulent in the disk (bottom-left panel Fig.~\ref{meridionaldiskB5}), and form a large-scale
helical structure along the BH spin axis (bottom panel).

\section{Summary and Future work}
\label{sec:summaryandfuturework}

In conclusion, if the NS tidally disrupts in a BHNS merger, its
internal magnetic fields are wound by the fluid motion into a
predominantly toroidal configuration. However, even very weak residual
poloidal fields in the disk may be exponentially amplified by MRI, generating
turbulence that drives accretion, ultimately launching jets and
providing the engine for an SGRB. 

Replicating this complete scenario with fully general relativistic MHD
simulations has been impossible, since numerically resolving MRI in
disks with such weak residual poloidal fields has been
computationally unfeasible. In our earlier work, equatorial symmetry was imposed,
which prevents poloidal fluid motion across the orbital plane,
thus suppressing any poloidal magnetic fields that could be
generated from this motion. 

In this work, we no longer impose this symmetry and, in addition,
choose asymmetric initial magnetic field configurations in the NS,
tilting its poloidal fields with respect to the orbital angular
momentum. Such asymmetries may be common, given the magnetic field
misalignment of PSR J0737-3039. We find that the more the fields are
tilted, the stronger the poloidal fields in the disk. When the initial
magnetic fields are tilted 90$^{\circ}$, the poloidal fields in the
remnant disk are amplified by about a factor of 8, producing
significantly more favorable conditions for modeling MRI. Our
``Generalized Lorenz'' gauge condition prevents the spurious growth of
magnetic fields that may plague some AMR simulations that use a
simpler ``algebraic gauge''~\cite{UIUCEMGAUGEPAPER}. Adopting GL, we
find that MRI is not resolvable with our chosen resolution and initial
magnetized BHNS models, even when the initial fields are purely
poloidal and have high amplitude.

To demonstrate that BHNS mergers may form an SGRB
engine under the right conditions, we perform the first simulation
that generates an SGRB engine from a BHNS remnant disk with a purely
poloidal field, which is dynamically weak initially, but strong
enough to adequately resolve MRI. We then observe magnetic turbulence,
followed by the large-scale collimation of magnetic fields and mildly
relativistic outflows perpendicular to the disk, sufficient to power
an SGRB.

We conclude that strong dipole poloidal fields in the remnant disks
from BHNS mergers can give rise to MRI and launch
Poynting-dominated outflows. Forming poloidal fields strong
enough to resolve MRI self-consistently through GRMHD simulations will
likely require asymmetric initial conditions.

Future work will focus on exploring new regions of parameter space
that may further increase the poloidal fields in the remnant disk and
the chances for launching jets. First, we intend to perform
simulations with more rapidly spinning BHs. With initial spin
parameter 0.75, the BH completely swallows the strongly-magnetized NS
core during merger, leaving only the weakly-magnetized outer layers in
the disk. A faster spinning BH may result in a larger, more strongly
magnetized disk. In addition, we plan to explore asymmetric
configurations in which the BH spin is misaligned with the orbital
angular momentum. Plasma in the resulting warped disk will have much
stronger motion parallel to the BH spin axis, likely enhancing the
poloidal fields and the possibility of establishing BHNS mergers as
plausible SGRB progenitors through fully general relativistic MHD
simulations.

\acknowledgments

The authors wish to thank Y.~T.~Liu, B.~Farris, C.~F.~Gammie, and N.~Vlahakis
for useful discussions. We also thank 
our REU team, including 
G.~Colten,
M.~Jin,
A.~Kim,
D.~Kolschowsky,
B.~Taylor,
and F.~Walsh, for assistance in producing the 3D 
visualizations. These visualizations were
created using the ZIB Amira software package~\cite{Stalling:AmiraVDA-2005}, and we gratefully 
acknowledge ZIB for providing us a license
license.
This paper was supported in part by NSF Grants AST-1002667, and
PHY-0963136 as well as NASA Grant NNX11AE11G at the
University of Illinois at Urbana-Champaign. This work used XSEDE, which is supported by NSF grant number OCI-1053575.

\bibliography{paper}

\end{document}